\documentclass[english,preprint,aip,jcp,superscriptaddress,showpacs,floatfix]{revtex4-1}
\usepackage{amsmath}
\usepackage{graphicx}
\usepackage{makeidx}

\newcommand{\bra}[1]{\left \langle #1 \right \vert}
\newcommand{\ket}[1]{\left \vert #1 \right \rangle}

\begin{document}

\title{Nonadiabatic Dynamics in Open Quantum-Classical Systems: Forward-Backward Trajectory Solution}

\author{Chang-Yu Hsieh}
\affiliation{Chemical Physics Theory Group, Department of Chemistry, University of Toronto, Toronto, ON, M5S 3H6 Canada}

\author{Raymond Kapral}
\affiliation{Chemical Physics Theory Group, Department of Chemistry, University of Toronto, Toronto, ON, M5S 3H6 Canada}

\begin{abstract}
A new approximate solution to the quantum-classical Liouville equation is derived starting from the formal solution of this equation in forward-backward form. The time evolution of a mixed quantum-classical system described by this equation is obtained in a coherent state basis using the mapping representation, which expresses $N$ quantum degrees of freedom in a $2N$-dimensional phase space.  The solution yields a simple non-Hamiltonian dynamics in which a set of $N$ coherent state coordinates evolve in forward and backward trajectories while the bath coordinates evolve under the influence of the mean potential that depends on these forward and backward trajectories. It is shown that the solution satisfies the differential form of the quantum-classical Liouville equation exactly. Relations to other mixed quantum-classical and semi-classical schemes are discussed.
\end{abstract}

\maketitle

\section{Introduction} \label{sec:intro}

Nonadiabatic processes are at the core of many physical phenomena, including population transfer among electronic system states, quantum coherent evolution of a system interacting with environmental degrees of freedom, electron and proton transfer reactions in condensed phase and biological systems, among others. In investigating such phenomena one often focuses on certain quantum degrees of freedom whose dynamics is of primary interest. These may be the electronic degrees of freedom of a chromophore excited by radiation to prepare the initial state of the system, the exciton states of a light harvesting system, or even the electron or proton degrees of freedom involved in the transfer of these particles. In such cases we are led to consider how these quantum degrees of freedom interact with the environment in which they reside. Interactions with the environment can lead to the breakdown of the Born-Oppenheimer approximation and one must consider nonadiabatic dynamics in such open quantum systems.

A number of different approaches have been developed to describe nonadiabatic dynamics. These include mean-field and a variety of surface-hopping schemes\cite{tully90,0chap-tully98,webster94,prezhdo99,truhlar04,bedard05,fischer11}, methods based on semi-classical evaluations of path integral formulations of quantum mechanics~\cite{miller78,meyer79,sun98,stock97,muller98,thoss99,thompson99,miller01,thoss04,stock05,bonella05,dunkel08,huo11} and descriptions based on the quantum-classical Liouville equation~\cite{kapral06}. An important ingredient in any approach dealing with nonadiabatic dynamics is the manner in which quantum coherence and decoherence are taken into account in the dynamics. The description of nonadiabtic dynamics necessarily entails dealing with coherence that is generated and destroyed as the system evolves while interacting with its environment. Many of the various nonadiabatic approaches that have been constructed deal with the issue of decoherence in various ways~\cite{hammesschiffer94,bittner95,subotnik11,shenvi11}.

Another characteristic of nonadiabatic schemes is the manner in which the environment is modeled. At the simplest level, the environment may be treated as a stochastic bath, which leads to reduced descriptions that do not explicitly include the environmental degrees of freedom in the evolution. Their effect only appears in certain parameters and terms that characterize the coupling to the environment. Schemes of this type include various quantum master equations~\cite{weiss99}, the Lindblad equation~\cite{lindblad76} and the Redfield and Bloch equations~\cite{redfield65,blum81}. Other methods explicitly account for the environmental degrees of freedom. It is challenging to treat large and complex systems fully quantum mechanically, although there are developments along these lines~\cite{meyer90,beck00,marques04,alon07,wang09,casida12}. Some methods, for example, some path integral methods, begin with a full quantum treatment and then make semi-classical approximations to obtain tractable solutions~\cite{stock05,bonella05,dunkel08,bukhman09,huo11}. Often the environment in which the quantum dynamics of interest occurs can be described by classical dynamics to a high degree of accuracy and this has spawned a number of mixed quantum-classical descriptions of nonadiabatic dynamics. Many surface-hopping schemes fall in this category as do some approximations to semi-classical path integral formulations and mean-field methods~\cite{0chap-tully98a,herman94,stock05}. Here we focus on descriptions based on the quantum-classical Liouville equation (QCLE).

The QCLE employs a partial Wigner representation of the environmental (bath) degrees of freedom and may be derived from full quantum dynamics by truncating the quantum evolution operator to first order in a small parameter related to the ratio of the characteristic masses of quantum and bath degrees of freedom~\cite{kapral99}. It may also be derived from partially linearized path integral formulations~\cite{shi04a,bonella10}, indicating the close connection between these different starting points. This equation has been shown to provide an accurate description of nonadiabtic dynamics in many applications and to account for quantum decoherence~\cite{kapral06}. A number of different methods, whose structure depends on the basis chosen to represent the quantum degrees of freedom, have been devised for its simulation~\cite{martens96,donoso98,mackernan02,mackernan08,horenko02,wan00,wan02}. Simulation methods that utilize an adiabatic basis can be cast into the form of surface-hopping dynamics, but in a way that includes coherent evolution segments that account for creation and destruction of coherence in a proper manner. More recently, as in some semi-classical approaches~\cite{stock05}, the mapping basis~\cite{chap-schwinger65} was used to describe the quantum degrees of freedom in the QCLE in a continuous classical-like manner, leading to a trajectory description in the full system phase space~\cite{kim-map08,nassimi10,kelly12}.

In this paper we also utilize the mapping representation but instead of dealing directly with the solution of the QCLE using a Liouville propagator, we start with its solution in terms of forward-backward quantum-classical propagators constructed some time ago~\cite{nielsen01}. With this starting point and the introduction of a coherent state basis~\cite{glauber63} we are able to obtain a solution of the QCLE that involves forward-backward trajectories of the coherent state variables, coupled to the evolution of the bath phase space variables.  Formally, both forward and backward trajectories are propagated forward in time.  The two sets of trajectories are distinguished and named by their association with the forward and backward quantum-classical propagators, respectively. This formulation leads to a simple set of non-Hamiltonian equations that describe the nonadiabatic dynamics of the system.

The outline of the paper is as follows: In Sec.~\ref{sec:QCLE} we sketch the important features of the QCLE, its representation in the mapping basis and formal solution in forward-backward form needed for our calculation. The forward-backward trajectory solution is constructed in Sec.~\ref{sec:FBsoln}, which contains the most important results of the paper. A discussion of the results is presented in Sec.~\ref{sec:disc}, while the Appendices give additional technical details of the calculation.

\section{Quantum-classical Liouville equation}\label{sec:QCLE}
We consider a quantum subsystem coupled to a bath. We assume that the dynamics of such a system is described by the quantum-classical Liouville equation~\cite{kapral99,alek81,geras82,boucher88,zhang88,mukamel94,martens96,horenko02}. For a quantum operator $\hat{B}_W(X)$, which depends on the classical phase space variables $X=(R,P)=(R_1,R_2,...,R_{N_b},P_1,P_2,...,P_{N_b})$ of the bath, this evolution equation takes the form,
\begin{equation} \label{eq:qcl-Bobs}
\frac{d}{d t}\hat{B}_W (X, t) =i  \hat{{\mathcal L}}\hat{B}_W (X,t),
\end{equation}
where the quantum-classical Liouville operator is
\begin{equation}\label{eq:qclop-abs}
i\hat{{\mathcal L}} \cdot = \frac{i}{\hbar}[\hat{H}_W,\cdot] - \frac{1}{2}(\{\hat{H}_W,\cdot\}
-\{\cdot,\hat{H}_W\}).
\end{equation}
Here the subscript $W$ refers to a partial Wigner transform over the bath degrees of freedom (DOF), $\hat{H}_W(X)$ is the partial Wigner transform of the total Hamiltonian of the system, $[\cdot,\cdot]$ is the commutator and $\{\cdot,\cdot\}$ is the Poisson bracket in the phase space of the classical variables $X$. The total Hamiltonian may be written as the sum of bath, subsystem and coupling terms,
\begin{equation} \label{eq:ham}
\hat{H}_W(X)= H_b(X)+\hat{h}_s+\hat{V}_c(R) \equiv H_b(X)+\hat{h}(R),
\end{equation}
where $H_b(X)=P^2/2M+V_b(R)$ is the bath Hamiltonian
with $V_b(R)$ the bath potential energy, $\hat{h}_s=\hat{p}^2/2m+\hat{V}_s$ is the
subsystem Hamiltonian with $\hat{p}$ and $\hat{V}_s$ the subsystem momentum and
potential energy operators, and $\hat{V}_c(R)$ is the coupling potential
energy operator. The masses of the subsystem and bath particles are $m$ and $M$, respectively.
The evolution equation for the density matrix $\hat{\rho}_W(X,t)$ is analogous to Eq.~(\ref{eq:qcl-Bobs}) with a change in sign of the evolution operator.

\subsection{Formal solution}
The QCLE may also be written in a form that is analogous to the quantum Liouville equation~\cite{nielsen01}:
\begin{equation} \label{eq:QCLEscript}
\frac{d}{d t}\hat{B}_W (X, t)=\frac{i}{\hbar}\Big(
\stackrel{\rightarrow}{{\mathcal H}}_{\Lambda}\hat{B}_W-\hat{B}_W
\stackrel{\leftarrow}{{\mathcal H}}_{\Lambda}\Big),
\end{equation}
where operators $\stackrel{\rightarrow}{{\mathcal H}}_{\Lambda}$ and $\stackrel{\leftarrow}{{\mathcal H}}_{\Lambda}$ are given by
\begin{eqnarray}\label{eq:sah}
{\stackrel{\rightarrow}{{\cal H}}_{\Lambda}}=\hat{H}_W\left(1+ \frac{\hbar \Lambda}{2i}\right), \quad
{\stackrel{\leftarrow}{{\cal H}}_{\Lambda}}= \left(1+ \frac{\hbar \Lambda}{2i}\right)\hat{H}_W,
\end{eqnarray}
with $\Lambda$ the negative of the Poisson bracket operator, $\Lambda = \stackrel{\leftarrow}{\nabla}_P \cdot
\stackrel{\rightarrow}{\nabla}_R-\stackrel{\leftarrow}{\nabla}_R \cdot
\stackrel{\rightarrow}{\nabla}_P$.

The formal solution of the QCLE can be expressed in either of two forms as
\begin{eqnarray}\label{bformalsol}
\hat{B}_W(X,t)&=&e^{i\hat{{\cal L}}t} \hat{B}_W(X) \\
&=& {\mathcal S}
\left(
e^{i{\stackrel{\rightarrow}{{\cal H}}_{\Lambda}}t/\hbar}\hat{B}_W(X)
e^{-i{\stackrel{\leftarrow}{{\cal H}}_{\Lambda}}t/\hbar} \right).\nonumber
\end{eqnarray}
The first equality follows from the formal solution of Eq.~(\ref{eq:qcl-Bobs}) while the second equality follows from Eq.~(\ref{eq:QCLEscript}). The ${\mathcal S}$ in this latter form simply specifies the order in which products of the left and right operators act in order to be identical with the first from involving the QCL operator. In particular, a general term ${\mathcal S}\left((\stackrel{\rightarrow}{{\mathcal H}}_{\Lambda})^j
\hat{B}_W (\stackrel{\leftarrow}{{\mathcal H}}_{\Lambda})^k\right)$ in the expansion of the exponential operators is composed of $\frac{(j+k)!}{j!k!}$ separate terms each with a prefactor of $\frac{j!k!}{(j+k)!}$. Each of these separate terms corresponds to a specific order in which the
$\stackrel{\rightarrow}{{\cal H}}_{\Lambda}$ and $\stackrel{\leftarrow}{{\cal H}}_{\Lambda}$ operators act on $\hat{B}_W$. This formal solution will be used in the calculations presented below.

\subsection{Mapping representation}
We will be concerned with the representation of the QCLE in the quantum subsystem basis and its equivalent representation in the mapping basis. The subsystem basis, $\{|\lambda\rangle ; \lambda =1, \dots, N\}$, is defined by the eigenvalue problem $\hat{h}_s |\lambda\rangle =\epsilon_\lambda |\lambda\rangle$, and a matrix element of an operator $\hat{B}_W(X)$ is given by ${B}^{\lambda \lambda'}_W (X)= \langle \lambda |\hat{B}_W(X)|\lambda'\rangle$.

The $|\lambda\rangle$ eigenfunctions of an $N$-state quantum subsystem can be replaced with eigenfunctions of $N$ fictitious harmonic oscillators~\cite{chap-schwinger65,stock05}, $|m_{\lambda}\rangle$, having occupation numbers which are limited to 0 or 1: $|\lambda\rangle\rightarrow|m_{\lambda}\rangle=|0_{1}, \cdots,1_{\lambda},\cdots0_{N}\rangle$. Creation and annihilation operators on these states, $\hat{a}_{\lambda}^{\dag}$ and $\hat{a}_{\lambda}$, respectively, are defined as
\begin{equation}
\label{eq:mappingg_creation_annihilation}
\hat{a}^{\dag}_\lambda  =  \frac{1}{\sqrt{2\hbar}} \left( \hat{q}_\lambda
-i\hat{p}_\lambda \right), \quad
\hat{a}_\lambda  =  \frac{1}{\sqrt{2\hbar}} \left( \hat{q}_\lambda +
i\hat{p}_\lambda \right),
\end{equation}
and satisfy the commutation relation $[ \hat{a}_\lambda, \hat{a}^{\dag}_{\lambda'}]   =  \delta_{\lambda,\lambda'}$. The actions of these operators on the single-excitation mapping states are $\hat{a}^{\dag}_\lambda \ket{0}  =  \ket{m_\lambda}$ and $\hat{a}_\lambda \ket{m_\lambda}  =  \ket{0}$, where
$\ket{0} = \ket{0_1 \dots 0_{N}}$ is the ground state of the mapping basis.

We may then define mapping versions of operators, $\hat{B}_{m}(X)$, given by
\begin{equation}
\hat{B}_{m}(X)=B_{W}^{\lambda\lambda'}(X)
\hat{a}_{\lambda}^{\dag}\hat{a}_{\lambda'},
\label{Bm_sub}
\end{equation}
so that a matrix element of $\hat{B}_W$ in the subsystem basis is equal to the matrix element of the corresponding mapping operator in the mapping single-excitation basis:
$B_{W}^{\lambda\lambda'}(X) =\langle\lambda|\hat{B}_{W}(X)|\lambda'\rangle=\langle
m_{\lambda}|\hat{B}_{m}(X)|m_{\lambda'}\rangle$. (The Einstein summation convention will used throughout although sometimes sums will be explicitly written if there is the possibility of confusion.)  In particular, the mapping Hamiltonian operator is
\begin{equation}
\label{eq:mapping_hamiltonian}
\hat{H}_m= H_b(X)+  h^{\lambda \lambda'}(R) \hat{a}^{\dag}_\lambda \hat{a}_{\lambda'}\equiv H_b(X)+\hat{h}_m,
\end{equation}
where we applied the mapping transformation only on the part of the Hamiltonian that involves
the subsystem DOF in Eq.~(\ref{eq:mapping_hamiltonian}). The pure bath term, $\hat{H}_{b}(X)$ in
Eq.~(\ref{eq:ham}), acts as an identity operator in the subsystem basis and is mapped onto the
identity operator of the mapping space.\\

The QCLE (\ref{eq:QCLEscript}) may now be written in terms of mapping operators as
\begin{equation} \label{eq:QCLEscript-map}
\frac{d}{d t}\hat{B}_m (X, t)=\frac{i}{\hbar}\Big(
\stackrel{\rightarrow}{{\mathcal H}^m_{\Lambda}}\hat{B}_m-\hat{B}_m
\stackrel{\leftarrow}{{\mathcal H}^m_{\Lambda}}\Big),
\end{equation}
where $\stackrel{\rightarrow}{{\mathcal H}^m_{\Lambda}}$ is given by ${\stackrel{\rightarrow}{{\cal H}^m_{\Lambda}}}=\hat{H}_m(1+ \hbar \Lambda/2i)$,
with an analogous definition for ${\stackrel{\leftarrow}{{\cal H}^m_{\Lambda}}}$. One may verify that the mapping space matrix elements of this equation are identical to the subsystem matrix elements of Eq.~(\ref{eq:QCLEscript}). Consequently, the formal solution of this equation is similar to that in Eq.~(\ref{bformalsol}) and is given by
\begin{equation}\label{bformalsol-map}
\hat{B}_m(X,t)={\mathcal S} \left(
e^{i{\stackrel{\rightarrow}{{\cal H}^m_{\Lambda}}}t/\hbar}\hat{B}_m(X)
e^{-i{\stackrel{\leftarrow}{{\cal H}^m_{\Lambda}}}t/\hbar} \right).
\end{equation}
This equation will form the starting point for the explicit solution of the QCLE in terms of forward-backward trajectories.

\section{Forward-backward trajectory solution}\label{sec:FBsoln}

The formal solution of the QCLE can be written in terms of a sequence of $M$ short-time propagators acting on the initial value of the operator:
\begin{equation}
\hat{B}_W(X,t)=e^{i\hat{{\cal L}} \Delta t_1} e^{i\hat{{\cal L}} \Delta t_2} \dots e^{i\hat{{\cal L}}\Delta t_M}\hat{B}_W(X),
\end{equation}
where $\Delta t_j=t_j-t_{j-1}=\tau$ for all $j$ with $t_0=0$ and $t_M=t$. (When information about a specific time step is needed we use the $\Delta t_j$ notation, otherwise the common value $\tau$ will be used.) Consequently, in view of Eq.~(\ref{bformalsol-map}), the formal solution applies in each time segment so $\hat{B}_W(X,t)$ may also be written as
\begin{eqnarray}\label{eq:mqc_soln-for}
&&\hat{B}_m (X,t) = {\mathcal S}\Big(e^{i\Delta t_1\stackrel{\rightarrow}{{\cal H}^m_{\Lambda}}/{\hbar}}
   {\mathcal S} \Big(e^{i \Delta t_2\stackrel{\rightarrow}{{\cal H}^m_{\Lambda}}/{\hbar}}  \dots \nonumber \\
&& \qquad {\mathcal S}\Big( e^{i \Delta t_M\stackrel{\rightarrow}{{\cal H}^m_{\Lambda}}/{\hbar}} \hat{B}_m(X)e^{-i \Delta t_M \stackrel{\leftarrow}{{\cal H}^m_{\Lambda}}/{\hbar}}\Big) \nonumber \\
&& \qquad \dots
e^{-i \Delta t_2 \stackrel{\leftarrow}{{\cal H}^m_{\Lambda}}/{\hbar}}\Big)
e^{-i \Delta t_1 \stackrel{\leftarrow}{{\cal H}^m_{\Lambda}}/{\hbar}}\Big),
\end{eqnarray}
where there are $M$ concatenated ${\mathcal S}\left( \cdots \right)$
brackets.

\subsection{Representation in coherent states}
In order to proceed with the evaluation we must consider the computation of the forward and backward propagators in this expression. To order $\tau^2$ we have
\begin{equation}
e^{i\tau\stackrel{\rightarrow}{{\cal H}^m_{\Lambda}}/{\hbar}}=  e^{\hat{H}_m \Lambda \tau/2} e^{i\hat{H}_m \tau/\hbar}+{\mathcal O}(\tau^2).
\end{equation}
Also, to order $\tau^2$ we may write the first exponential operator as
\begin{eqnarray}
\label{eq:coherent_approx0}
&&e^{\hat{H}_m\Lambda \tau/2} = 1 +
\frac{\tau}{2}\hat{H}_m\Lambda + \dots,
\\
&& \quad =
1 +\frac{\tau}{2} H_b(X) \Lambda+\frac{\tau}{2}
h^{\lambda \lambda'}\hat{a}^{\dag}_\lambda\hat{a}_{\lambda'}\Lambda +
\dots,
\nonumber \\
&& \quad =
1 + \frac{\tau}{2} H_b(X) \Lambda +
\frac{\tau}{2} \left(h^{\lambda \lambda'}\hat{a}_{\lambda'}\hat{a}^{\dag}_{\lambda} -
\mathrm{Tr_s}\;h \right)\Lambda+ \dots, \nonumber
\end{eqnarray}
where we have reversed the normal-ordered product of annihilation and creation operators into an anti-normal
order form using their commutation relation. The by-product of reversing the ordering of creation and
annihilation operators is the emergence of a trace term in the last line of this equation. Since the trace term is independent of the quantum state, it may be combined with the bath potential, $V_{0}(R)=V_b(R)-\mathrm{Tr_s}\;h(R)$, to give $H_{0}(X)=P^2/2M+V_{0}(R)$ so that we have the simpler form of Eq.~(\ref{eq:coherent_approx0}),
\begin{equation}\label{eq:coherent_approx}
e^{\hat{H}_m\Lambda \tau/2} =
1 + \frac{\tau}{2} H_{0}(X) \Lambda +
\frac{\tau}{2} h^{\lambda \lambda'}\hat{a}_{\lambda'}\hat{a}^{\dag}_{\lambda} \Lambda+
{\mathcal O}(\tau^2).
\end{equation}
In this form, the propagator can be expressed conveniently in coherent states~\cite{glauber63}.

We define the coherent states $\ket{z}$ in the mapping space,
\begin{equation}
\label{eq:coherent_state}
\hat{a}_{\lambda} \ket{z}= z_\lambda \ket{z},\quad \bra{z}\hat{a}_{\lambda}^\dagger= z_\lambda^* \bra{z},
\end{equation}
where $\ket{z}$ is a coherent state with $N$ degrees of freedom and the eigenvalue is $z_\lambda= (q_\lambda + i  p_\lambda )/\sqrt{2\hbar}$. The variables $q=(q_1, \dots, q_{N})$ and $p=(p_1, \dots, p_{N})$ are mean coordinates and momenta of the harmonic oscillators in the state $\ket{z}$, respectively; i.e.,
we have $\bra{z} \hat{q}_\lambda \ket{z} = q_\lambda$ and $\bra{z} \hat{p}_\lambda \ket{z} = p_\lambda$.

The coherent states form an overcomplete basis; thus, we have to specify the inner product between any pair of coherent states and the
resolution of identity~\cite{glauber63}. The inner product is
\begin{eqnarray}
\label{eq:coherent_state_properties}
\bra{z}z' \rangle & = & e^{-\frac{1}{2}(|z|^2+|z'|^2) +z^*\cdot z'}\nonumber \\
 &  = & e^{-\frac{1}{2}(|z-z'|^2) -i\Im (z\cdot z^{\prime *})}.
\end{eqnarray}
The norm of the inner product measures how far away the two coherent states $\ket{z}$ and $\ket{z'}$ are in the
phase space of coherent state variables. The resolution of the identity is
\begin{equation}
1= \int \frac{d^2z}{\pi^{N}} \ket{z}\bra{z},
\end{equation}
where $d^2z=d(\Re(z)) d(\Im(z))=dq dp/(2\hbar)^N$.

Given these properties of the coherent states, we may insert the resolution of the identity in the bath Hamiltonian terms and between the $\hat{a}_{\lambda'}$ and $\hat{a}^{\dag}_{\lambda}$ operators in Eq.~(\ref{eq:coherent_approx}) to obtain
\begin{eqnarray}
\label{eq:coherent_op_diagonal_form}
e^{\frac{\tau}{2}\hat{H}_m\Lambda} & = &
(1 + \frac{\tau}{2}H_{0}(X)\Lambda )\int \frac{d^2z}{\pi^{N}}\ket{z}\bra{z}\nonumber \\
&&+
\frac{\tau}{2} \int \frac{d^2z}{\pi^{N}}
h^{\lambda \lambda'}\hat{a}_{\lambda'}\ket{z}\bra{z}\hat{a}^{\dag}_{\lambda}\Lambda + {\mathcal O}(\tau^2) \nonumber \\
& = &
\int \frac{d^2z}{\pi^{N}}\ket{z}\Big(1 + \frac{\tau}{2}(H_{0}(X) +h^{\lambda \lambda'}z^*_\lambda z_{\lambda'}) \Lambda\nonumber \\
&&
\qquad \qquad + {\mathcal O}(\tau^2)\Big)\bra{z} \nonumber \\
& =&
\int \frac{d^2z}{\pi^{N}}\ket{z}
e^{\frac{\tau}{2}{H}_{cl}(X,z)\Lambda} \bra{z} + {\mathcal O}(\tau^2).
\end{eqnarray}
In this calculation we used Eq.~(\ref{eq:coherent_state}) to eliminate the annihilation and creation operators in Eq.~(\ref{eq:coherent_op_diagonal_form}).
Note that $h^{\lambda \lambda'}z^*_\lambda z_{\lambda'}=
\frac{1}{2\hbar}h^{\lambda \lambda'}(q_{\lambda'}q_{\lambda} + p_{\lambda}p_{\lambda'})$ since $h^{\lambda \lambda'}$ is symmetric. In the last line of Eq.~(\ref{eq:coherent_op_diagonal_form}) we defined the ``classical" Hamiltonian
\begin{eqnarray}\label{eq:Hcl-def}
{H}_{cl}(X,z) &=& H_{0}(X) + h^{\lambda \lambda'}z^*_\lambda z_{\lambda'} \equiv H_{0}(X) + h_{cl}(R,z)\nonumber \\
&=& \frac{P^2}{2M}+ h_{s,cl}(z) +V_{cl}(R,z),
\end{eqnarray}
where $V_{cl}(R,z)=V_0(R)+V_c^{\lambda \lambda'}(R)z^*_\lambda z_{\lambda'}$.

The operator ${H}_{cl}(X,z)\Lambda$ acts on all bath phase space variables to its right. Therefore, it is convenient to introduce a notation that makes this action evident. More specifically, we let
\begin{equation}
\label{eq:Hcl}
{H}_{cl}(X,z)\Lambda= \frac{\partial H_{cl}}{\partial P}\cdot \frac{\stackrel{\rightarrow}{\partial} }{\partial R}-\frac{\partial H_{cl}}{\partial R}\cdot \frac{\stackrel{\rightarrow}{\partial} }{\partial P} \equiv i\stackrel{\rightarrow}{{\mathcal L}}(X,z),
\end{equation}
so that
\begin{equation}
\label{eq:opf}
e^{\frac{\tau}{2}\hat{H}_m\Lambda} =\int \frac{d^2z}{\pi^{N}}\ket{z}
e^{i\stackrel{\rightarrow}{{\mathcal L}}(X,z) \tau/2} \bra{z} + {\mathcal O}(\tau^2).
\end{equation}
Similarly we can define
\begin{equation}
\Lambda{H}_{cl}(X,z)= \frac{\stackrel{\leftarrow}{\partial} }{\partial P} \cdot \frac{\partial H_{cl}}{\partial R}-\frac{\stackrel{\leftarrow}{\partial} }{\partial R} \cdot \frac{\partial H_{cl}}{\partial P} \equiv -i\stackrel{\leftarrow}{{\mathcal L}}(X,z),
\end{equation}
and
\begin{equation}
\label{eq:opb}
e^{-\frac{\tau}{2}\Lambda\hat{H}_m} = \int \frac{d^2z}{\pi^{N}}\ket{z}
e^{i\stackrel{\leftarrow}{{\mathcal L}}(X,z) \tau/2} \bra{z} + {\mathcal O}(\tau^2).
\end{equation}

The other quantity that will enter in the evaluation of the time evolution is the action of the exponential operator $e^{i\hat{H}_m(X)\tau/\hbar}$ on a coherent state. In Appendices A and B we show that
\begin{eqnarray}
\label{eq:coherent_state_evolution}
e^{-i\hat{H}_m(X)\tau/\hbar}\ket{z} & = &
e^{-iH_b(X)\tau/\hbar}e^{-i\hat{h}_m(R)\tau/\hbar}\ket{z},
\nonumber \\
& = &
e^{-iH_b(X)\tau/\hbar}\ket{z(\tau)},
\end{eqnarray}
with $z(\tau)$ determined from the solution of the evolution equation,
\begin{eqnarray}
\label{eq:coherent_state_dynamics}
\frac{dz_{\lambda}}{dt} =-\frac{i}{\hbar}\frac{\partial h_{cl}}{\partial{z}^*_{\lambda}}.
\end{eqnarray}

\subsection{Time evolution of an operator}

These results may now be used to compute the value of the matrix elements of an operator $\hat{B}_W(X,t)$ in the subsystem basis: $B^{\lambda \lambda'}_W(X,t)=\langle m_\lambda | \hat{B}_m(X,t) |m_{\lambda'} \rangle$. We have
\begin{eqnarray}\label{eq:mqc_soln-mat}
&&{B}^{\lambda \lambda'}_W (X,t) =\sum_{\mu \mu'}\int \prod_{i=1}^M \frac{d^2z_i}{\pi^{N}} \frac{d^2z'_i}{\pi^{N}}\nonumber \\
&& \qquad \quad \langle m_{\lambda} \ket{z_1}{\mathcal S}\Big(
e^{i\stackrel{\rightarrow}{{\mathcal L}}(X,z_1) \frac{\Delta t_1}{2}} \bra{z_1} e^{i\hat{H}_m \frac{\Delta t_1}{\hbar}}\ket{z_2}\nonumber \\
&& \qquad \quad
   {\mathcal S} \Big(
e^{i\stackrel{\rightarrow}{{\mathcal L}}(X,z_2) \frac{\Delta t_2}{2}} \bra{z_2} e^{i\hat{H}_m \frac{\Delta t_2}{\hbar}}  \dots \ket{z_M}\nonumber \\
&& \qquad \quad {\mathcal S}\Big(
e^{i\stackrel{\rightarrow}{{\mathcal L}}(X,z_M) \frac{\Delta t_M}{2}} \bra{z_M} e^{i\hat{H}_m \frac{\Delta t_M}{\hbar}} |m_{\mu} \rangle \nonumber \\
&& \qquad \quad {B}^{\mu \mu'}_W(X)\langle m_{\mu'}|
 e^{-i\hat{H}_m \frac{\Delta t_M}{\hbar}}\ket{z^\prime_M} e^{i\stackrel{\leftarrow}{{\mathcal L}}(X,z^\prime_M) \frac{\Delta t_M}{2}}\Big) \nonumber \\
&& \qquad \quad \bra{z^\prime_M}  \dots
 e^{-i\hat{H}_m \frac{\Delta t_2}{\hbar}}\ket{z^\prime_2} e^{i\stackrel{\leftarrow}{{\mathcal L}}(X,z^\prime_2) \frac{\Delta t_2}{2}}\Big)
 \nonumber \\
&& \qquad \quad  \bra{z^\prime_2}
 e^{-i\hat{H}_m \frac{\Delta t_1}{\hbar}}\ket{z^\prime_1} e^{i\stackrel{\leftarrow}{{\mathcal L}}(X,z^\prime_1) \frac{\Delta t_1}{2}} \Big)\bra{z^\prime_1} m_{\lambda'}\rangle,
\end{eqnarray}

We may now make use of the definition of the ${\mathcal S}$ operator to rewrite the actions of the right and left operators acting on the bath coordinates of an arbitrary operator $\hat{A}_W(X)$ in terms of a single effective operator ${\mathcal L}_{e}(X,z,z')$ that depends on the coherent state variables $z$ and $z'$ associated with the forward and backward propagators, respectively. In Appendix C we show that
\begin{eqnarray} \label{eq:effL-prop}
&&{\mathcal S} \Big(e^{i\stackrel{\rightarrow}{{\mathcal L}}(X,z) \frac{\tau}{2}} \hat{A}_W(X) e^{i\stackrel{\leftarrow}{{\mathcal L}}(X,z^\prime) \frac{\tau}{2}}\Big) \\
&& \qquad \qquad = e^{i {\mathcal L}_{e}(X,z,z') \tau}\hat{A}_W(X) \equiv \hat{A}_W(X_\tau).\nonumber
\end{eqnarray}
The explicit form of $i {\mathcal L}_{e}(X,z,z')$ is
\begin{eqnarray} \label{eq:effL}
i {\mathcal L}_{e}(X,z,z')=\frac{P}{M}  \cdot \frac{\partial}{\partial R} -\frac{\partial V_e(X,z,z')}{\partial R}  \cdot \frac{\partial}{\partial P},
\end{eqnarray}
where $V_e(X,z,z')=(V_{cl}(R,z)+V_{cl}(R,z'))/2$. From Eqs.~(\ref{eq:effL-prop}) and (\ref{eq:effL}) we can see that the time evolution of the bath coordinates under the effective Liouville operator is given by the solutions of the equations
\begin{equation}
\frac{dR}{dt}=\frac{P}{M}, \qquad \frac{dP}{dt}=-\frac{\partial V_e(X,z,z')}{\partial R}.
\end{equation}

These results may be used in the expression for $B^{\lambda \lambda'}_W (X,t)$ in Eq.~(\ref{eq:mqc_soln-mat}) to give
\begin{eqnarray}\label{eq:mqc_soln-mat2}
&&{B}^{\lambda \lambda'}_W (X,t) =\sum_{\mu \mu'}\int \prod_{i=1}^M \frac{d^2z_i}{\pi^{N}} \frac{d^2z'_i}{\pi^{N}}
\langle m_{\lambda} \ket{z_1} \bra{z^\prime_1} m_{\lambda'}\rangle \nonumber \\
&& \qquad \quad
e^{i{\mathcal L}_e(X,z_1,z_1') \frac{\Delta t_1}{2}} \Big( \bra{z_1} e^{i\hat{H}_m \frac{\Delta t_1}{\hbar}}\ket{z_2}\nonumber \\
&& \qquad \quad
    e^{i{\mathcal L}_e(X,z_2,z_2') \frac{\Delta t_2}{2}} \Big( \bra{z_2} e^{i\hat{H}_m \frac{\Delta t_2}{\hbar}}  \dots \ket{z_M}\nonumber \\
&& \qquad \quad
e^{i{\mathcal L}_e(X,z_M,z_M') \frac{\Delta t_M}{2}} \Big( \bra{z_M} e^{i\hat{H}_m \frac{\Delta t_M}{\hbar}} |m_{\mu} \rangle \nonumber \\
&& \qquad \quad {B}^{\mu \mu'}_W(X)\langle m_{\mu'}|
 e^{-i\hat{H}_m \frac{\Delta t_M}{\hbar}}\ket{z^\prime_M} \Big)  \\
&& \qquad \quad \bra{z^\prime_M}  \dots
 e^{-i\hat{H}_m \frac{\Delta t_2}{\hbar}}\ket{z^\prime_2} \Big)
\bra{z^\prime_2}  e^{-i\hat{H}_m \frac{\Delta t_1}{\hbar}}\ket{z^\prime_1}  \Big).\nonumber
\end{eqnarray}
This expression can be evaluated by applying the operators from left to right. For example, the action of the first effective bath operator updates the bath phase space coordinates from $X=X_{t_0}$ to $X_{t_1}$. Thus,
\begin{eqnarray}\label{eq:mqc_soln-mat3a}
&&{B}^{\lambda \lambda'}_W (X,t) =\sum_{\mu \mu'}\int \prod_{i=1}^M \frac{d^2z_i}{\pi^{N}} \frac{d^2z'_i}{\pi^{N}}
\langle m_{\lambda} \ket{z_1} \bra{z^\prime_1} m_{\lambda'}\rangle \nonumber \\
&& \qquad
 \Big( \bra{z_1} e^{i\hat{H}_m(X_{t_1}) \frac{\Delta t_1}{\hbar}}\ket{z_2}\nonumber \\
&& \qquad \dots {B}^{\mu \mu'}_W(X_{t_1}) \dots
\bra{z^\prime_2}  e^{-i\hat{H}_m(X_{t_1}) \frac{\Delta t_1}{\hbar}}\ket{z^\prime_1}  \Big),
\end{eqnarray}
The coherent state matrix elements can now be evaluated using Eq.~(\ref{eq:coherent_state_evolution}) to give
\begin{eqnarray}\label{eq:mqc_soln-mat3}
&&{B}^{\lambda \lambda'}_W (X,t) =\sum_{\mu \mu'}\int \prod_{i=1}^M \frac{d^2z_i}{\pi^{N}} \frac{d^2z'_i}{\pi^{N}}
\langle m_{\lambda} \ket{z_1} \bra{z^\prime_1} m_{\lambda'}\rangle \nonumber \\
&& \qquad
 \Big(e^{iH_b(X_{t_1})\Delta t_1/\hbar} \langle z_1(t_1) \ket{z_2} e^{i{\mathcal L}_e(X_{t_1},z_2,z_2') \frac{\Delta t_2}{2}} \Big(\bra{z_2}\dots \nonumber \\
 && \qquad \qquad {B}^{\mu \mu'}_W(X_{t_1}) \dots \ket{z^\prime_2} \Big)
e^{-iH_b(X_{t_1})\Delta t_1/\hbar} \bra{z^\prime_2} z^\prime_1(t_1)\rangle  \Big) \nonumber \\
&& \qquad
 =\sum_{\mu \mu'}\int \frac{d^2z_i}{\pi^{N}} \frac{d^2z'_i}{\pi^{N}}
\langle m_{\lambda} \ket{z_1} \bra{z^\prime_1} m_{\lambda'}\rangle \nonumber \\
&& \qquad \qquad
 \Big( \langle z_1(t_1) \ket{z_2} e^{i{\mathcal L}_e(X_{t_1},z_2,z_2') \frac{\Delta t_2}{2}} \Big(\bra{z_2}\dots \nonumber \\
&& \qquad  \qquad \qquad {B}^{\mu \mu'}_W(X_{t_1}) \dots
\ket{z^\prime_2} \Big) \bra{z^\prime_2} z^\prime_1(t_1)\rangle  \Big) .
\end{eqnarray}
In writing the last equality we canceled the phase factors involving $H_b(X_{t_1})$.

At this point we can see how a description involving continuous trajectories may be constructed. The classical bath propagator for the next time step from $t_1$ to $t_2$, $e^{i{\mathcal L}_e(X_{t_1},z_2,z_2') \frac{\Delta t_2}{2}}$, involves the coherent state phase space variables $z_2$ and $z_2'$ which may take any values from the set of coherent state values. The coherent states involved in the matrix elements $\langle z_1(t_1) \ket{z_2}$ and $\bra{z^\prime_2}z^\prime_1(t_1)\rangle$ are not orthogonal since the coherent states are overcomplete. However, in view of Eq.~(\ref{eq:coherent_state_properties}), we see that the overlap between two coherent states decays rapidly if their phase space coordinates differ significantly. Consequently we assume that $\langle z_1(t_1) \ket{z_2} \approx \pi^{N}\delta(z_2-z_1(t_1))$ and $\bra{z^\prime_2}z^\prime_1(t_1)\rangle \approx \pi^{N}\delta(z_2^\prime-z_1^\prime(t_1))$. Then performing the integrals over $z_2$ and $z_2^\prime$ we obtain
\begin{eqnarray}\label{eq:mqc_soln-mat4}
&&{B}^{\lambda \lambda'}_W (X,t) =\sum_{\mu \mu'}\int  \frac{d^2z_1}{\pi^{N}} \frac{d^2z'_1}{\pi^{N}}
\langle m_{\lambda} \ket{z_1} \bra{z^\prime_1} m_{\lambda'}\rangle  \\
&& \qquad \qquad \int \prod_{i=3}^M \frac{d^2z_i}{\pi^{N}} \frac{d^2z'_i}{\pi^{N}}
 \Big(  e^{i{\mathcal L}_e(X_{t_1},z_1(t_1),z_1'(t_1)) \frac{\Delta t_2}{2}} \nonumber \\
&& \qquad  \qquad \qquad \times \Big(\bra{z_1(t_1)}\dots {B}^{\mu \mu'}_W(X_{t_1}) \dots
\ket{z^\prime_1(t_1)} \Big)   \Big) .\nonumber
\end{eqnarray}
All coherent state and bath phase space variables have now been updated to time $t_1$ and process can now be repeated for all $M$ time steps, starting with the application of the effective bath evolution operator for the time step $\Delta t_2$. The result of this process is the simple expression
\begin{eqnarray}\label{eq:mqc_soln-mat5}
&&{B}^{\lambda \lambda'}_W (X,t) =\sum_{\mu \mu'}\int \frac{d^2 z_1}{\pi^{N}} \frac{d^2 z^{\prime}_1}{\pi^{N}}
\langle m_{\lambda} \ket{z_1} \bra{z^\prime_1} m_{\lambda'}\rangle  \\
&& \qquad  \qquad \qquad \times \Big(\bra{z_1(t)} m_\mu \rangle {B}^{\mu \mu'}_W(X_{t})
\langle m_{\mu'}\ket{z^\prime_1(t)} \Big)   \Big) .\nonumber
\end{eqnarray}
The matrix elements between coherent states and the single-excitation mapping states may be evaluated explicitly to give
\begin{equation}
\langle m_\lambda \ket{z}= z_\lambda e^{-|z|^2/2}.
\end{equation}
Writing this expression in terms of the $x=(q,p)$ variables, and using the fact that $\sum_\nu (q_\nu^2+p_\nu^2)$ is conserved under coherent state dynamics, we obtain
\begin{eqnarray}\label{eq:mqc_soln-mat6}
&&{B}^{\lambda \lambda'}_W (X,t) =\sum_{\mu \mu'}\int dx dx^{\prime} \phi(x) \phi(x') \nonumber \\
&& \qquad \times \frac{1}{2\hbar} (q_{\lambda}+ip_{\lambda})(q'_{\lambda^\prime}-ip'_{\lambda^\prime}) {B}^{\mu \mu'}_W(X_{t})\nonumber \\
&& \qquad \times \frac{1}{2\hbar} (q_{\mu}(t)-ip_{\mu}(t))(q'_{\mu^\prime}(t)+ip'_{\mu^\prime}(t)),
\end{eqnarray}
where $\phi(x)=\left(2\pi\hbar\right)^{-N}e^{-\sum_\nu (q_\nu^2+p_\nu^2)/2\hbar}$ is the normalized Gaussian distribution function and we have removed the subscript $1$ from the  dummy coherent state variables. The coupled equations of motion governing this evolution are
\begin{eqnarray}\label{eq:vol-nonH}
&&\frac{d q_\lambda}{dt}= \frac{\partial H_{cl}(R,P,q,p)}{\partial p_\lambda}, \quad \frac{d p_\lambda}{dt}= -\frac{\partial H_{cl}(R,P,q,p)}{\partial q_\lambda} \nonumber \\
&&\frac{d q'_\lambda}{dt}= \frac{\partial H_{cl}(R,P,q',p')}{\partial p'_\lambda}, \quad \frac{d p'_\lambda}{dt}= -\frac{\partial H_{cl}(R,P,q',p')}{\partial q'_\lambda}\nonumber \\
&&\frac{dR}{dt}=\frac{P}{M}, \qquad  \frac{dP}{dt}=-\frac{\partial H_e(R,P,q,p,q',p')}{\partial R},
\end{eqnarray}
where
\begin{eqnarray}\label{eq:Heff}
&&H_e(R,P,q,p,q',p')=\\
&& \qquad \qquad \frac{1}{2}( H_{cl}(R,P,q,p)+H_{cl}(R,P,q',p')).\nonumber
\end{eqnarray}
Equation~(\ref{eq:mqc_soln-mat6}) and the associated evolution equations~(\ref{eq:vol-nonH}) are the results we set out to derive. They constitute a simple algorithm for obtaining a solution to the QCLE.  Figure~\ref{fig1} presents a schematic picture that depicts the dynamics of coordinates prescribed by the evolution equations~(\ref{eq:vol-nonH}). As noted earlier, although both forward and backward trajectories are propagated forward in time, the two sets of trajectories arise from the forward and backward quantum-classical propagators, respectively.

Earlier it was shown that the solution to the QCLE in the mapping basis can be given in terms of an ensemble of entangled trajectories~\cite{kelly12}. The solution in Eq.~(\ref{eq:vol-nonH}) is consistent with this interpretation in that the forward and backward trajectories of the coherent state variables are linked by the evolution of the bath variables and the evolution equations are in non-Hamiltonian form. A more detailed link between these two different approaches to the QCLE in the mapping basis is a topic that merits further study.


\subsection{Back to Differential Form}\label{sec:diff_form}

In this section we show that the solution constructed above is indeed a solution of the QCLE. We do this by deriving the QCLE in the subsystem basis by constructing a finite-difference expression for the time evolution of $B^{\lambda\lambda'}_W(X,t)$.
We first write the matrix element for $\bra{\lambda} \hat{B}_W(X,t+\tau)\ket{\lambda'}$ using Eq.~(\ref{eq:mqc_soln-mat5}),
\begin{eqnarray}
\label{eq:first_order_rho_expansion}
&&B^{\lambda,\lambda'}_W(X,t+\tau) =
\sum_{\mu \mu'}\int d^2z(t) d^2z'(t) \phi(z)\phi(z')
\\
& & \quad \times
z_{\lambda}(t) {z}^{\prime*}_{\lambda'}(t) {z}^*_{\mu}(t+\tau) z^{\prime}_{\mu'}(t+\tau) \left(e^{i{\mathcal L}_e\tau}B^{\mu\mu'}_W(X,t)\right),\nonumber
\end{eqnarray}
where $\phi(z)=\pi^{-N}e^{-|z|^2/2}$. We then expand to first order in $\tau$ to obtain
\begin{eqnarray} \label{eq:tau-expansion}
&&B^{\lambda,\lambda'}_W(X,t+\tau) \approx \sum_{\mu \mu'}\int d^2z(t) d^2z'(t) \phi(z)\phi(z')
\nonumber\\
& & \qquad \qquad \times z_{\lambda}(t) {z}^{\prime*}_{\lambda'}(t) \Big[ {z}^*_{\mu}(t) z^\prime_{\mu'}(t) B^{\mu\mu'}_W(X,t) \nonumber \\
&& \qquad  \qquad+\tau \Big(  {z}^*_{\mu}(t)\frac{\partial z^\prime_{\mu'}}{\partial t} + z^\prime_{\mu'}(t)\frac{\partial {z}^*_{\mu}}{\partial t} \Big)B^{\mu\mu'}_W(X,t) \nonumber \\
&& \qquad \qquad +
\tau {z}^*_{\mu}(t) z^\prime_{\mu'}(t) i{\mathcal L}_e  B^{\mu\mu'}_W(X,t)  \Big] + {\mathcal O}(\tau^2).
\end{eqnarray}
The integrals over $z(t)$ and $z'(t)$ may be performed and, after rearranging terms and taking the limit $\tau \to 0$, the result is (some details are given in Appendix D),
\begin{eqnarray}
\label{eq:derived_mqcle}
&&\lim_{\tau \to 0} \frac{ B^{\lambda\lambda'}_W(t+\tau) -B^{\lambda\lambda'}_W(t)}{\tau} = \frac{ d }{dt}B^{\lambda\lambda'}_W(t)
\\
&& \quad =
\bra{\lambda} \frac{i}{\hbar} \left[ \hat{H}_{W},
\hat{B}_W \right] \ket{\lambda'}\nonumber \\
&&\quad -\frac{1}{2}\left(\bra{\lambda} \left\{ \hat{H}_W, \hat{B}_W
  \right\} - \left\{ \hat{B}_W, \hat{H}_W   \right\} \ket{\lambda'}\right) , \nonumber
\end{eqnarray}
which is the QCLE.

The QCLE in the subsystem basis is a first order differential equation with respect to time; therefore, it only describes how the matrix elements of $\hat{B}_W(X,t)$ at the beginning and the end of a time step are related.  That our solution is found to satisfy the QCLE is consistent with
the fact that all approximations used to derive the evolution in a single time step are exact to $\mathcal O (\tau^2)$.
However, in order to connect the trajectories of coherent state phase variables from adjacent time steps, we made the approximation, $\bra{z_i(\tau)} z_{i+1} \rangle \approx \pi^N \delta(z_{i+1}-z_i(\tau))$.
To understand the effects of this approximation, we consider how our solution would be modified if the approximation were not made.  One way to re-formulate the solution is to insert a set of single-excitation mapping states between every inner product of coherent states, i.e. $\bra{z_i(\tau)} z_{i+1} \rangle =  \sum_{\mu_i} \bra{z_i(\tau)} m_{\mu_i} \rangle \bra{m_{\mu_i}} z_{i+1} \rangle$.  Once the mapping states are inserted, one loses the continuous trajectory picture in the coherent state phase space but one can formally integrate out the $z_i$ and $z'_i$ variables in sequential (or chronological) order.
This sequence of formal integrations is equivalent to evaluations of the matrix elements of $\hat{B}_W(X,t)$ at every time step.  Computationally, this is a very demanding task because one needs to sample, propagate and integrate out coherent state trajectories at every time step.  However, this prescription (a continuous evolution of matrix elements) coincides exactly with the dynamics one would expect from the QCLE in the subsystem basis.

At this point, it is obvious that the coherent-state orthogonality approximation replaces the continuous evolution of the matrix elements, $B^{\lambda\lambda'}_W(X,t)$, with continuous trajectories, $z(t)$ and $z'(t)$. Instead of taking $B^{\mu\mu'}_W(X,t-\tau)$ as the starting point to compute $B^{\lambda\lambda'}_W(X,t)$  at the next time step. The orthogonality approximation actually takes the operator $\ket{z(t-\tau)}\bra{z(0)} \hat{B}_m(X,0) \ket{z'(0)} \bra{z'(t-\tau)}$ as the starting point and further propagates trajectories from the previous time step to obtain $\ket{z(t)}\bra{z(0)} \hat{B}_m(X,0) \ket{z'(0)} \bra{z'(t)}$.
Although the orthogonality approximation inevitably yields nonlocal errors, it does provide a computationally efficient way to simulate the dynamics.  Other semi-classical approaches for solving the system-bath dynamics indicate that this is a sensible approximation to make.  For instance, if we do not use the orthogonality approximation then we can write our solution in the form of a standard coherent state path integral.  Application of the stationary phase approximation will yield the same set of equations of motion for the coherent state phase variables. Similar coherent state dynamics was obtained in the context of a different semi-classical framework~\cite{huo11}.

Finally, we comment on the fact that the semiclassical analysis yields exact quantum mechanical solution for quadratic Hamiltonians.  This is certainly true when the system is isolated from the bath.  The same also holds true for our solution; if there are no bath terms then there is no need to make the orthogonality approximation.  However, when a bath is present, the semi-classical analysis is equivalent to implicitly making the orthogonality approximation, which becomes exact in the limit $\hbar \rightarrow 0$ in view of Eq.~(\ref{eq:coherent_state_properties}). The potential source of errors, which arises from the system-bath interactions, can easily be overlooked because it is eliminated as soon as semi-classical conditions are imposed.

\section{Discussion}\label{sec:disc}

The results derived above provide a simple simulation algorithm for the dynamics described by the QCLE. Most often it is the average value of an operator (or correlation function) that is of interest. The average value of a quantum operator $\hat{B}_W(X,t)$ is given by
\begin{eqnarray}\label{eq:Bsub}
\overline{B(t)}&=&\int dX\; {\rm Tr} \; (\hat{B}_{W}(X)\hat{\rho}_{W}(X,t))\\
&=& \int dX\; B_{W}^{\lambda\lambda'}(X,t)\rho_{W}^{\lambda'\lambda}(X),\nonumber
\end{eqnarray}
where the trace is taken in the quantum subsystem space. Using Eq.~(\ref{eq:mqc_soln-mat5}) for the time evolution of $B_{W}^{\lambda\lambda'}(X,t)$, the average value may be computed by sampling over the coherent state variables and initial density matrix element $\rho_{W}^{\lambda'\lambda}(X)$.

Our solution for $B_{W}^{\lambda\lambda'}(X,t)$ has a number of elements in common with other approaches that have been devised to simulate nonadiabatic dynamics and it is instructive to make comparisons with methods that have been constructed in a similar spirit.

\subsection{Comparison with partially linearized path integral methods}

First, we  draw comparisons between two mixed quantum-classical formalisms: the QCLE and partially linearized path integral methods.
The formal equivalence between the two formalisms was established in a general setting~\cite{bonella10} when the subsystem DOF are expressed as quantum operators.  Therefore, the close resemblance between our solution and that of Huo and Coker~\cite{huo11} is expected, since they are approximate solutions to the QCLE and a particular form of the partially linearized path integral, respectively.  However, in view of the derivation of our solution presented above, the result  in Ref.~[\onlinecite{huo11}] is not an exact solution of QCLE. In our formalism, $H_{cl}$ defined in Eq.~(\ref{eq:Hcl-def}) contains $V_0(R)=V_b(R)-{\rm Tr}_s \hat{h}$ instead of simply the bath potential $V_b(R)$. Recall that the trace term arose from the commutation relation for the annihilation and creation operators and the need to use an anti-normal order for the product of these operators to evaluate the short-time propagator. If this trace term is absent one can show that the solution does not satisfy the differential form of the QCLE.

The system Hamiltonian, $\hat{H}_W(X) = H_b(X) +\hat{h}(R)$, can be written in an equivalent form $\hat{H}_W(X) = H_b(X) + (Tr_s\hat{h}(R))/N + \hat{\bar{h}}(R)$, where $\hat{\bar{h}}(R)$ is traceless. Since this is an identity, the QCLE is independent of the choice of the form which is used in this equation. Our solution is also independent of the way the Hamiltonian is written, although the equations of motion take a somewhat different form. If the Hamiltonian with the trace removed is used in the derivation, the evolution equations have the same structure as is in Eq.~(\ref{eq:vol-nonH}) but $H_{cl}(X,z)$ in Eq.~(\ref{eq:Hcl-def}) is replaced by ${H}_{cl}(X,z) = H_{0}(X) + \bar{h}^{\lambda \lambda'}z^*_\lambda z_{\lambda'}$ with $H_{0}(X) \to H_b(X)+(Tr_s\hat{h}(R))/N$.~\cite{footnote:H-tr-rem} When the calculation given in Sec.~\ref{sec:diff_form} is repeated with this form of the Hamiltonian the QCLE is again obtained, confirming that the different but equivalent forms of the Hamiltonian yield the same evolution.

However, this is not the case when other approximate theories are considered. In particular, it was shown~\cite{kelly12} that the choice of Hamiltonian form is crucial in the Poisson Bracket Mapping Equation (PBME) approximation to the QCLE (discussed below). When the traceless form is used, dynamical instabilities that arise in the course of the evolution can be tamed, while if the original form of the Hamiltonian is used the instabilities can lead to difficulties.

The form of the Hamiltonian also affects the nature of the dynamics in the semi-classical approach used in Ref.~[\onlinecite{huo11}]. While the evolution equations in this approach differ from those in Eq.~(\ref{eq:vol-nonH}), the equivalence is restored between the two solutions if the traceless form of the Hamiltonian is used. The reason that the partially linearized path integral solution depends sensitively on the form of the Hamiltonian is due to the semi-classical approach used to solve the dynamics. According to the semi-classical calculation, the dynamics of the bath momenta are governed by the force, $-\frac{1}{2} \left( \partial \tilde{H}_{cl}(X,z) / \partial R + \partial \tilde{H}_{cl}(X,z') / \partial R\right)$, where $\tilde{H}_{cl}(X,z) = H_b(X) + h_{cl}(R,z)$.  The Hamiltonian $\tilde{H}_{cl}(X,z)$ misses the term $-Tr_s \hat h(R)$ in ${H}_{cl}(X,z)$ in Eq.~(\ref{eq:Hcl-def}) in the current formulation. This extra term is required to restore the equivalence between the solution using the original Hamiltonian and that using the traceless form of the Hamiltonian.

\subsection{Comparison with Poisson bracket mapping equation}

Next, we compare the current solution to the PBME approximation to the quantum-classical Liouville equation~\cite{kim-map08,nassimi10,kelly12}, which is obtained from the mapping form of the QCLE by dropping an excess coupling term~\cite{nassimi10}.  In the case of an isolated subsystem, one can perform a change of variables $\bar{z} = (z + z')/2$ and $\Delta z = z - z'$ and show that both the mean, $\bar{z}$, and the difference, $\Delta z$, variables follow exactly the same Hamiltonian dynamics, as described in Eq.~(\ref{eq:coherent_state_dynamics}) with no $R$ dependence.  This implies that if $\bar{z}(0) = \Delta z(0)$ then $\bar{z} = \Delta z(t)$ for all $t$.  Since the computation of the time evolution of an operator in the subsystem basis requires integration over the entire coherent state phase space, as prescribed in Eq.~(\ref{eq:mqc_soln-mat6}), $\Delta z$ becomes a redundant variable.  A direct comparison between the two methods can be made if one either integrates out $\Delta z$ or replaces the integral of $\Delta z$ by integral of $\bar z$ as follows,
\begin{eqnarray}
\label{eq:diffvar_trick}
&& \int \frac{d^2 \bar z}{\pi^N} e^{-2\vert \bar z \vert^2} \bar z_{\lambda} \bar z_{\mu}(t) \int \frac{d^2 \Delta z}{\pi^N} e^{-\frac{1}{2}\vert \Delta z \vert^2} \Delta z_{\mu'}(t) \Delta z_{\lambda'} =
\nonumber \\
&&\qquad 2^N \int \frac{d^2 \bar z}{\pi^N} e^{-2\vert \bar z \vert^2}  \left(\bar z_{\lambda}\bar z^*_{\mu}(t)\bar z_{\mu'}(t)\bar z^*_{\lambda'} - \bar z_{\lambda}\bar z^*_{\lambda'}\delta_{\mu,\mu'} \right.
\nonumber  \\
&&\qquad \left.  + \frac{1}{4} \delta_{\lambda,\lambda'}\delta_{\mu,\mu'}\right).
\end{eqnarray}
The above identity can be easily proved in a basis that diagonalizes the Hamiltonian, followed by transformation of the resulting identity back to the original basis, in the same spirit as the computation of the exact coherent state dynamics in Appendix A.

After properly removing $\Delta z$, one can show that Eq.~(\ref{eq:mqc_soln-mat6}) reduces to
\begin{eqnarray}
\label{eq:pbme_form}
&& B^{\lambda\lambda'}_W(X,t) =
\sum_{\mu,\mu'}\int dx
\left(
\bar z_\lambda \bar{z}^*_{\lambda'}
\bar z_\mu^*(t) \bar z_{\mu'}(t) -
\frac{1}{4}
\bar z_{\lambda} \bar{z}^*_{\lambda'}\delta_{\mu,\mu'}
\right. \nonumber \\
&& \qquad
\left.
-\frac{1}{2} \bar z_\mu(t) \bar z^*_{\mu'}(t)
\delta_{\lambda,\lambda'} + \frac{1}{8} \delta_{\lambda,\lambda'}
\delta_{\mu,\mu'}
\right) 4\phi(x) B^{\mu\mu'}_W(X,t),
\nonumber \\
&& \qquad
= \sum_{\mu,\mu'} \int dx g_{\lambda,\lambda'}(x) B^{\mu\mu'}_W(X,t) c_{\mu\mu'}(x(t)),
\end{eqnarray}
where $x = (q = \sqrt{2\hbar}\Re \bar z , p = \sqrt{2\hbar} \Im \bar z)$ and the functions~\cite{kelly12} $g_{\lambda\lambda'}(x) \equiv (\ket{m_\lambda}\bra{m_{\lambda'}})_W$ and $c_{\mu\mu'}(x) \equiv (\hat a^{\dag}_\lambda \hat a_{\lambda'})_W$ represent the Wigner transformation $(.)_W$ of the outer product of states and a pair of annihilation and creation operators, respectively.  The last expression in Eq.~(\ref{eq:pbme_form}) is exactly the evolution of $B^{\lambda\lambda'}_W(X,t)$ in the PBME method.  Furthermore, the Wigner transformation variables, $x$, in the PBME method follow the same Hamiltonian dynamics derived above for the mean coordinates of the coherent state variables. Despite the very different starting points of the two solutions,  this comparison reveals the close relation between the dynamics of Wigner transformed coordinates and the mean coordinates in the coherent state phase space.  Although, this close relation can only be made obvious after the effects of difference variables are properly taken into account of and removed (either explicitly integrated out or replaced using Eq.~(\ref{eq:diffvar_trick})). Essentially, the dynamical information encoded in the coherent states variables of $2N$ harmonic oscillators can be merged and be encoded in $N$ Wigner transformed coordinates.

We next comment on the comparison to the PBME method in the presence of a bath. One may linearize the bath potential $V_e(X, z, z') \approx (V_{cl}(R,\bar z ) + \frac{\partial V_{cl}}{\partial \bar z}\Delta z + V_{cl}(R,\bar z) -  \frac{\partial V_{cl}}{\partial \bar z}\Delta z )/2 = V_{cl}(R,\bar z)$ such that the dynamics of the bath variables only depends on the mean coordinates $\bar z$.  The bath potential linearization allows one to properly remove the difference variables and encode the approximate dynamics in $N$ harmonic oscillators.  Repeating the same calculations and using the coherent-state orthogonality approximation, one can show that Eq.~(\ref{eq:pbme_form}) still holds in the general mixed quantum-classical setting.

\subsection{Comparison with semi-classical schemes}

Finally, we compare our results to some semi-classical schemes.
The mapping representation consolidates the way subsystem dynamics is handled in the QCLE and in some semi-classical schemes. For instance, in the case of a linearized bath potential, the Hamiltonian dynamics prescribed by our solution is also identical to that in the semi-classical path integral approach of Stock and Thoss~\cite{stock05,thoss99} as well as the linearized semi-classical initial value representation (LSC-IVR) of Miller~\cite{miller78,miller01,miller07}. Furthermore, the full version of the current solution also handles the subsystem dynamics in ways similar to the forward-backward semi-classical initial value representation (FB-IVR) approaches\cite{thompson99,sun99,thoss01} that uses the Herman-Kluk propagator.  One difference is that the forward and backward trajectories are not linked in the present solution.

Finally, we observe that the classical-like system-bath dynamics prescribed in our solution could be similar to that of a mixed semi-classical scheme~\cite{sun97} in which the bath DOF and subsystem DOF are treated with LSC-IVR and the FB-IVR, respectively. Further investigations into the subtle connections between the our solution of the QCLE and other semi-classical schemes might inspire further developments in nonadiabatic quantum dynamics.

\begin{acknowledgments}
This work was supported in part by a grant from the Natural Sciences and Engineering Council of Canada.
\end{acknowledgments}

\appendix
\section*{Appendix A: Exact evolution of coherent states}

We restrict this analysis to real-valued, symmetric, quadratic Hamiltonian operators, $\hat{h}_m$, which are the only type of Hamiltonian encountered in the mapping formalism. It is always possible to diagonalize such a Hamiltonian matrix, to obtain
\begin{eqnarray}
\label{eq:symmetricH}
\hat{h}_m & = &
\sum_{\lambda,\lambda'}h^{\lambda\lambda'}\hat{a}^{\dag}_{\lambda}\hat{a}_{\lambda'}=\sum_{\lambda,\lambda'}\sum_{\mu}M_{\lambda
  \mu}h^d_{\mu}M^T_{\mu \lambda'}
\hat{a}^{\dag}_{\lambda}\hat{a}_{\lambda'},
\nonumber \\
& = & \sum_{\mu} h^d_{\mu}
\left(\sum_{\lambda}\hat{a}^{\dag}_{\lambda}M_{\lambda \mu}\right)
\left(\sum_{\lambda'}M^T_{\mu \lambda}\hat{a}_{\lambda'}\right)
\nonumber \\
& = &\sum_{\mu} h^d_{\mu} \hat{b}^{\dag}_{\mu} \hat{b}_{\mu} \equiv \hat{h}^d_m,
\end{eqnarray}
where the operators $\hat{b}^{\dag}_{\mu}$ and $\hat{b}_{\mu}$ are defined
in the second line of the equation.  We use the superscript $d$ to
emphasize that the Hamiltonian is now put in the diagonal form with
respect to operators $\hat{b}_{\mu}$ and $\hat{b}^{\dag}_{\mu}$.  Since the Hamiltonian is real and symmetric, the matrix $M$ is an orthogonal matrix.

With respect to the new operators $\hat{b}_{\mu}$ and $\hat{b}^{\dag}_{\mu}$, we define the coherent state
$\ket{y}$ by
\begin{eqnarray}
\label{eq:new_coherent}
\hat{b}_{\mu}\ket{y} =y_{\mu} \ket{y}, \quad
\bra{y}\hat{b}^{\dag}_{\mu} = \bra{y} y^*_{\mu},
\end{eqnarray}
where $y_{\mu}=\frac{1}{\sqrt{2\hbar}}\left(\tilde{q}_\mu+i\tilde{p}_\mu\right)$.

Consider time evolution of the coherent state $\ket{y}$ with $N$ degrees of freedom,
\begin{eqnarray}
\label{eq:diagonalized_propagator}
e^{-\frac{i}{\hbar}\hat{h}^d_m t}\ket{y} & = &
e^{-\frac{i}{\hbar}\hat{h}^d_m t}
\otimes_{\nu=1}^N \left\{e^{-\vert y_\nu
  \vert^2}\sum_{m=0}^{\infty}\frac{y_\nu^m}{\sqrt{m!}}\ket{m}_\nu\right\},
\nonumber \\
& = & \otimes_{\nu=1}^N \left\{ e^{-\vert y_\nu  \vert^2}\sum_{m=0}^{\infty}\frac{(y_ie^{-\frac{\nu}{\hbar}h^d_{\nu}t})^m}{\sqrt{m!}}\ket{m}_\nu
\right\},  \nonumber \\
& = & \ket{y(t)},
\end{eqnarray}
where $y_\nu(t) = y_\nu(0)e^{-\frac{ih^d_{\nu}}{\hbar}t}$. In this calculation we used the expansion of a coherent state in terms of a complete set of harmonic oscillator states:
\begin{equation}
\label{eq:coh-state-expansion}
\ket{y_\nu} =
e^{-\vert y_\nu \vert^2}\sum_{m=0}^{\infty}\frac{y_\nu^m}{\sqrt{m!}}\ket{m}_\nu.
\end{equation}
Equation~(\ref{eq:diagonalized_propagator}) implies the equation of motion,
\begin{equation}
\frac{dy_\nu}{dt} = -\frac{i}{\hbar} h^d_{\nu} y_\nu(t)=-\frac{i}{\hbar}\frac{\partial \tilde{h}^d_{cl}}{\partial y^*_{\nu}},
\end{equation}
where $\tilde{h}^d_{cl} = \sum_\mu h^d_{\mu}y^*_\mu y_\mu$.
If we substitute in the variables $\tilde{q}$ and $\tilde{p}$ into the
equation of motion for $y$ then we get the usual Hamilton's equation for $\tilde{q}$ and $\tilde{p}$.

Next, we prove that Hamilton's equation is invariant under the linear transformation, $y_\mu = \sum_{\lambda}M_{\mu\lambda}^Tz_\lambda$, and
$y^*_\mu = \sum_{\lambda}z^*_\lambda M_{\lambda\mu}$. This proceeds as follows:
\begin{eqnarray}
\label{eq:original_z_eom}
\frac{dz_{\lambda}}{dt} & = &
\sum_{\mu}M_{\lambda \mu}\frac{dy_\mu}{dt}=-
\frac{i}{\hbar}\sum_{\mu}M_{\lambda \mu}\frac{\partial \tilde{h}^d_{cl}}{\partial y^*_\mu}\nonumber
\\
& = &-\frac{i}{\hbar} \sum_{\mu}M_{\lambda \mu}h^d_{\mu}y_\mu=-\frac{i}{\hbar} \sum_{\mu,\lambda'}
M_{\lambda,\mu}h^d_{\mu}M^T_{\mu \lambda'}z_{\lambda'}
\nonumber \\
& = &
-\frac{i}{\hbar}\sum_{\lambda'}h^{\lambda \lambda'}z_{\lambda'}=
-\frac{i}{\hbar}\frac{\partial h_{cl}}{\partial z^*_{\lambda}}.
\end{eqnarray}

\section*{Appendix B: Matrix elements of the unitary evolution operator in the
  single excitation subspace}

In this Appendix we evaluate matrix elements of the form
$\bra{m_\lambda}e^{-\frac{it}{\hbar}\hat{h}_m}\ket{m_{\lambda'}}$, where
$\hat{h}_m$ is still the real-valued and symmetric Hamiltonian
considered in Appendix A. We evaluate this matrix element in two
ways: directly and also in terms of matrix elements
$\bra{\tilde{m}_{\lambda}}e^{-\frac{i}{\hbar}\hat{h}^d_m t}\ket{\tilde{m}_{\lambda'}}$
via a linear transformation.  The state
$\ket{\tilde{m}_{\lambda}} = \ket{0_1 \dots 1_{\tilde{\lambda}} \dots 0_N}$ is an N-harmonic-oscillator state with a single excitation on the $\lambda$-th
oscillator, $h_{\lambda}^d\hat{b}^{\dag}_{\lambda}\hat{b}_{\lambda}$.

First, we prove that $\ket{m_\lambda} = \sum_{\mu} M^T_{\mu \lambda}
\ket{\tilde{m}_{\mu}}$.  This is straightforward since $\ket{m_\lambda} =
\hat{a}^{\dag}_{\lambda}\ket{0}$ and $\ket{\tilde{m}_{\mu}}=\hat{b}^{\dag}_{\mu}\ket{0}$, where
$\ket{0}$ is the common ground state.  Therefore, the two states are
related by the orthogonal matrix $M$, which was used to establish the linear
transformation between $\hat{a}^{\dag}_\lambda$ and $\hat{b}^{\dag}_{\mu}$. The evaluation proceeds as follows:
\begin{eqnarray}
\label{eq:transition_elements}
&&\bra{m_{\lambda'}} e^{-\frac{it}{\hbar}\hat{h}_m} \ket{m_{\lambda}}
= \bra{\tilde{m}_{\mu'}}M_{\lambda' \mu'}
e^{-\frac{it}{\hbar}\hat{h}^d_m} M^T_{\mu \lambda} \ket{\tilde{m}_{\mu}}
\nonumber \\
&& \quad =  \int \frac{d\tilde{x}}{(2\pi\hbar)^{N}}
M_{\lambda'\mu'}M^T_{\mu \lambda}
\bra{\tilde{m}_{\mu'}}e^{-\frac{it}{\hbar}\hat{h}^d_m} \ket{y}
\bra{y}\tilde{m}_{\mu}\rangle \nonumber \\
&& \quad =  \int \frac{d\tilde{x}}{(2\pi\hbar)^{N}}
M_{\lambda'\mu'}M^T_{\mu \lambda}
\bra{\tilde{m}_{\mu'}}y(t) \rangle
\bra{y}\tilde{m}_{\mu}\rangle  \\
&& \quad =  \int \frac{d\tilde{x}}{(2\pi\hbar)^{N}}
M_{\lambda\mu}M^T_{\mu' \lambda'}
y_{\mu'}(t) y^*_{\mu}e^{-\frac{1}{2}\vert y(t) \vert^2}
e^{-\frac{1}{2}\vert y \vert^2}, \nonumber \\
&& \quad =   \int \frac{d\tilde{x}}{(2\pi\hbar)^{N}}
z_{\lambda'}(t) z^*_{\lambda}
e^{-\vert z \vert^2}
=   \int \frac{dx}{(2\pi\hbar)^{N}}
z_{\lambda'}(t) z^*_{\lambda}
e^{-\vert z \vert^2},\nonumber
\end{eqnarray}
where $d\tilde{x} = d\tilde{q} d\tilde{p}$ and $dx=dq dp$.
To obtain the above result we used the relation $z_{\mu} = M^T_{\nu\mu}y_{\nu}=M_{\mu\nu}y_{\nu}$ to re-express the $y$ variables in terms of $z$ variables and employed the volume element transformation, $d\tilde{x}=dx \left \vert \mathrm{det} \left[\partial y_\alpha/\partial z_\beta \right] \right\vert= dx \left \vert \mathrm{det} M \right\vert =dx$, since $\left \vert \mathrm{det} M \right\vert=1$.  Since the $y(t)$ variables satisfy Hamilton's equations, $\vert y(t) \vert ^2 = \vert y \vert^2$. Finally, we note that $\vert y \vert^2 = \sum_{\lambda} y^*_\lambda y_\lambda = \sum_{\lambda,\mu,\mu'} M^T_{\mu'\lambda}M_{\lambda\mu} z^*_{\mu'}z_{\mu} = \sum_{\mu} z^*_{\mu}z_{\mu}=\vert z \vert^2$, completing the results needed to obtain Eq.~({\ref{eq:transition_elements}).

Next, we compute $e^{-\frac{it}{\hbar}\hat{h}_m} \vert z \rangle$ directly, where $\ket{z}$ is defined by $\hat{a}_\lambda \ket{z} = z_\lambda \ket{z}$. To carry out this calculation we reconsider Eq.~(\ref{eq:transition_elements}),
\begin{eqnarray}
\label{eq:transition_elements2}
&&\bra{m_{\lambda'}} e^{-\frac{i}{\hbar}\hat{h}_m t} \ket{m_{\lambda}} =\int \frac{dx}{(2\pi\hbar)^N}
\bra{m_{\lambda'}}e^{-\frac{i}{\hbar}\hat{h}_m t}\ket{z}\langle z \ket{m_{\lambda}}
\nonumber \\
&& \qquad =
\int \frac{dx}{(2\pi\hbar)^N}
\bra{m_{\lambda'}}e^{-\frac{i}{\hbar}\hat{h}_m t}\ket{z}z^*_{\lambda}e^{-\frac{1}{2}\vert z \vert^2}.
\end{eqnarray}
Comparing the last lines of Eqs.~(\ref{eq:transition_elements}) and (\ref{eq:transition_elements2}), we see that
$\bra{m_{\lambda'}}e^{-\frac{i}{\hbar}\hat{h}_m t}\ket{z} =
z_{\lambda'}(t) e^{-\frac{1}{2}\vert z(t) \vert^2} =
\bra{m_{\lambda'}} z(t) \rangle$.  Since the identities hold for
all possible $\bra{m_{\lambda'}}$ and $\ket{z}$, we can identify
$e^{-\frac{i}{\hbar}\hat{h}_m t}\ket{z} = \ket{z(t)}$.

\section*{Appendix C: Effective Liouville Operator}

Below, we prove the identity in Eq.~(\ref{eq:effL-prop}) that relates the
forward and backward bath propagators to the effective Liouville operator:
\begin{eqnarray}
\label{eq:eff_liouville_proof}
&&{\mathcal S} \Big(e^{i\stackrel{\rightarrow}{{\mathcal L}}(X,x) \frac{\tau}{2}} \hat{A}_W(X) e^{i\stackrel{\leftarrow}{{\mathcal L}}(X,x^\prime) \frac{\tau}{2}}\Big) \nonumber \\
&& \quad=
\sum_{j=0}^{\infty}\sum_{k=0}^{j}\frac{(i\tau/2)^j}{(j-k)!k!}
S\left((\stackrel{\rightarrow}{{\mathcal L}})^{k}\hat{A}_W(\stackrel{\leftarrow}{{\mathcal L}'})^{j-k}\right)\nonumber
\\
&& \quad=
\sum_{j=0}^{\infty}\frac{(i\tau/2)^j}{j!}
\sum_{k=0}^{j}
\left(\begin{array}{c}
j \\ k
\end{array}\right)
S\left((\stackrel{\rightarrow}{{\mathcal L}})^{k}\hat{A}_W(\stackrel{\leftarrow}{{\mathcal L}'})^{j-k}\right)\nonumber \\
&&\quad =
\sum_{j=0}^{\infty}\frac{(i\tau/2)^j}{j!}
\sum_{k=0}^{j}
\sum_{\{p\}} \stackrel{\rightarrow}{{\mathcal L}}^{(p_1)} \stackrel{\rightarrow}{{\mathcal L}}^{(p_2)} \dots \stackrel{\rightarrow}{{\mathcal L}'}^{(p_j)}\hat{A}_W\nonumber \\
&& \quad  =
\sum_{j=0}^{\infty}\frac{(i\tau)^j}{j!}
\left(\frac{1}{2}(\stackrel{\rightarrow}{{\mathcal L}}+\stackrel{\rightarrow}{{\mathcal L}'})\right)^j\hat{A}_W(X)\nonumber \\
&& \quad=
e^{i {\mathcal L}_{e}(X,x,x') \tau}\hat{A}_W.
\end{eqnarray}
In these expressions we used the shorthand notations, $\stackrel{\rightarrow}{{\mathcal L}} = \stackrel{\rightarrow}{{\mathcal L}}(X,x)$ and $\stackrel{\leftarrow}{{\mathcal L}'} =\stackrel{\leftarrow}{{\mathcal L}}(X,x')$, and the definition of ${\mathcal S}$ in going from the third to fourth lines. The sum on $\{p\}$ denotes a sum over all permutations.

\section*{Appendix D: Differential Form}

The term zeroth order in $\tau$ in Eq.~(\ref{eq:tau-expansion}) is easily computed by performing the integrals over $z(t)$ and $z'(t)$ and the result is simply $B^{\mu\mu'}_W(X,t)$. The first term of order $\tau$, which we call $I_1$, involves time derivatives coherent state variables. Using the equations of motion for the coherent state variables and performing the integrals we find
\begin{eqnarray}
I_1&=& \frac{i}{\hbar} \bra{\lambda} [\hat{h},\hat{B}_W(X,t)]\ket{\lambda'} \nonumber \\
&=&\frac{i}{\hbar} \bra{\lambda} [\hat{H}_W,\hat{B}_W(X,t)]\ket{\lambda'},
\end{eqnarray}
which is the first term in the QCL operator in Eq.~(\ref{eq:qclop-abs}).

Next, we consider the first-order term involving the evolution of the spatial coordinates of the bath as given by the effective Liouville operator. Inserting its definition in Eq.~(\ref{eq:effL}), $i {\mathcal L}_{e}(X,z,z')=\frac{P}{M}  \cdot \frac{\partial}{\partial R} -\frac{\partial V_e(X,z,z')}{\partial R}  \cdot \frac{\partial}{\partial P}$, the evaluation of the term involving
$\frac{\partial B^{\mu\mu'}_W}{\partial R}\cdot\frac{P}{M}$ is straightforward since it does not contain the coherent state variables. Performing the integrals over these variables yields $I_2=\frac{\partial B^{\lambda \lambda'}_W}{\partial R}\cdot\frac{P}{M}$. The remaining terms require more attention since they involves the force acting on the bath variables, which depends on the effective potential where $V_e(X,z,z')=(V_{cl}(R,z)+V_{cl}(R,z'))/2$. Denoting this contribution $I_3$, we have
\begin{eqnarray}
\label{eq:1st_order_bath_momentum}
I_3 & = & - \sum_{\mu \mu'}\int d^2z(t) d^2z'(t) \phi(z)\phi(z')
\nonumber \\
& &
\times z_{\lambda}(t){z}^{\prime *}_{\lambda'}(t){z}^*_{\mu}(t)z'_{\mu'}(t)
\frac{\partial B^{\mu\mu'}_W}{\partial P} \cdot \frac{\partial
  {H}_{e}(R,z,z')}{\partial R} \nonumber \\
& = & - \sum_{\mu \mu'}\int d^2z(t) d^2z'(t) \phi(z)\phi(z')
\nonumber \\
& &
\times z_{\lambda}(t){z}^{\prime *}_{\lambda'}(t){z}^*_{\mu}(t)z'_{\mu'}(t)
\frac{\partial B^{\mu\mu'}_W}{\partial P} \cdot \nonumber \\
&& \left[\frac{\partial {V}_{0}(R)}{\partial R} +\frac{1}{2}
\frac{\partial {V}_{cl}(R,z)}{\partial R} +
\frac{1}{2}\frac{\partial {V}_{cl}(R,z')}{\partial R} \right]
\nonumber \\
& = &
-\frac{\partial B^{\lambda \lambda'}_W}{\partial P}\frac{\partial
{V}_0(R)}{\partial R}+I_{3_1} + I_{3_2},
\end{eqnarray}

The $I_{3_1}$ integral may be evaluated as follows:
\begin{eqnarray}
\label{eq:1st_order_force_unprimed_coherent_state}
I_{3_1} &= & -\frac{1}{2} \sum_{\mu \mu'}\int d^2z(t) d^2z'(t) \phi(z)\phi(z')\nonumber \\
&& \times z_{\lambda}(t){z}^{\prime *}_{\lambda'}(t){z}^*_{\mu}(t)z'_{\mu'}(t) \frac{\partial B^{\mu\mu'}_W}{\partial P}
\frac{\partial {V}_{cl}(R,z)}{\partial R}
\nonumber \\
& = &
-\frac{1}{4\hbar} \sum_{\mu \alpha \alpha'}\int d^2z(t) \phi(z)
z_{\lambda}(t){z}^*_{\mu}(t)  \nonumber \\
&& \frac{\partial B^{\mu \lambda'}_W}{\partial P}
\cdot \frac{\partial {V}_c^{\alpha \alpha'}(R,z)}{\partial   R} z_{\alpha}(t) {z}^*_{\alpha'}(t)
\nonumber \\
& = &
-  \frac{1}{4\hbar} \sum_{\mu \alpha \alpha'}\int d^2z(t) \phi(z) \frac{\partial B^{\mu\lambda'}_W}{\partial P}
\cdot \frac{\partial {V}_c^{\alpha \alpha'}(R,z)}{\partial R}\nonumber \\
&&  \left[
\vert z_{\lambda}(t) \vert ^2 \vert z_{\alpha}(t) \vert^2
\delta_{\alpha \alpha'}
\delta_{\mu\lambda}\left(1-\delta_{\alpha \lambda}\right)
\right.
\nonumber \\
& &
\left.
+ \vert z_\lambda(t) \vert ^2 \vert z_{\mu}(t) \vert ^2
\delta_{\alpha \lambda}\delta_{\alpha' \mu}\left(1-\delta_{\alpha \alpha'}\right) \right.\nonumber \\
&&\left.+ \vert z_{\lambda}(t) \vert ^4
\delta_{\alpha \alpha'}\delta_{\alpha \lambda}\delta_{\mu \lambda}
\right] \nonumber
\end{eqnarray}

Performing the $z$ integrals we find
\begin{eqnarray}
I_{3_1} & =  &
- \frac{1}{2} \Big(\sum_{\alpha \neq \lambda}\frac{\partial
   {V}_c^{\alpha \alpha}}{\partial R}\frac{\partial
  B_W^{\lambda \lambda'}}{\partial P}
 +
\sum_{\mu \ne \lambda} \frac{\partial
 {V}_c^{\lambda \mu}}{\partial R}\frac{\partial
  B_W^{\mu \lambda'}}{\partial P}
\nonumber \\
&&
+2 \frac{\partial
  {V}_c^{\lambda \lambda}}{\partial R} \frac{\partial B_W^{\lambda \lambda'}}{\partial P}
\Big) \nonumber \\
& =  &
-\frac{1}{2}  \bra{\lambda}
\Big(
 \frac{\partial  \hat{V}_c}{\partial R}
\frac{\partial \hat{B}_W}{\partial P} +
\frac{\partial  \mathrm{Tr}_s  \hat{V}_c}{\partial R}
\frac{\partial \hat{B}_W}{\partial P}
\Big) \ket{\lambda'}\nonumber\\
&=&
-\frac{1}{2}  \bra{\lambda}
\Big(
 \frac{\partial  \hat{h}}{\partial R}
\frac{\partial \hat{B}_W}{\partial P} +
\frac{\partial  \mathrm{Tr}_s  \hat{h}}{\partial R}
\frac{\partial \hat{B}_W}{\partial P}
\Big) \ket{\lambda'}.
\end{eqnarray}
In writing the last line of this equation we used the fact that the subsystem Hamiltonian is independent of $R$ so $\hat{V}_c$ and be replaced by $\hat{h}$.

Similarly, the $I_{3_2}$ integral can be evaluated to give,
\begin{eqnarray}
\label{eq:1st_order_force_primed_coherent_state}
I_{3_2} &= & -\frac{1}{2}  \bra{\lambda}
\left(
\frac{\partial \hat{B}_W}{\partial P}
\frac{\partial  \hat{h}}{\partial R}
+
\frac{\partial \hat{B}_W}{\partial P}
\frac{\partial  \mathrm{Tr}_s  \hat{h}}{\partial R}
\right) \ket{\lambda'}.
\end{eqnarray}
Recall that $V_0(R)=V_b(R)-{\rm Tr}_s \hat{h}$ so that $\frac{\partial B^{\lambda \lambda'}_W}{\partial P}\frac{\partial
{V}_0(R)}{\partial R}=\frac{\partial B^{\lambda \lambda'}_W}{\partial P}\frac{\partial
{V}_b(R)}{\partial R}-\frac{\partial B^{\lambda \lambda'}_W}{\partial P}\frac{\partial
\mathrm{Tr}_s  \hat{h}}{\partial R}$. Given these results the entire $I_3$ integral is
\begin{equation}
I_3= -\frac{1}{2}\left(\bra{\lambda} \left\{ \hat{H}_W, \hat{B}_W
  \right\}- \left\{ \hat{B}_W, \hat{H}_W  \right\} \ket{\lambda'}\right),
\end{equation}
where the ${\rm Tr}_s \hat{h}$ terms arising from the $V_0$, $I_{3_1}$ and $I_{3_2}$ canceled.


\begin{thebibliography}{64}
\bibitem [{foo()}]{footnote:H-tr-rem}%
  \BibitemOpen
  \href@noop {} {}\bibinfo {note} {The use of the traceless form of the
  Hamiltonian from the outset simplfies the calculation. In particular, when
  the anti-normal form of the annihilation and creation operator product is
  introduced to evaluate the propagator, the trace term does not appear and the
  stucture of the equations is simpler. We have chosen to carry out the
  derivation with the usual form of the Hamiltonian to stress that the final
  results are independent of how one choses to write the
  Hamiltonian.}\BibitemShut {Stop}%
\end{thebibliography}

%

\newpage


\begin{figure}
 \includegraphics[width=0.8\linewidth]{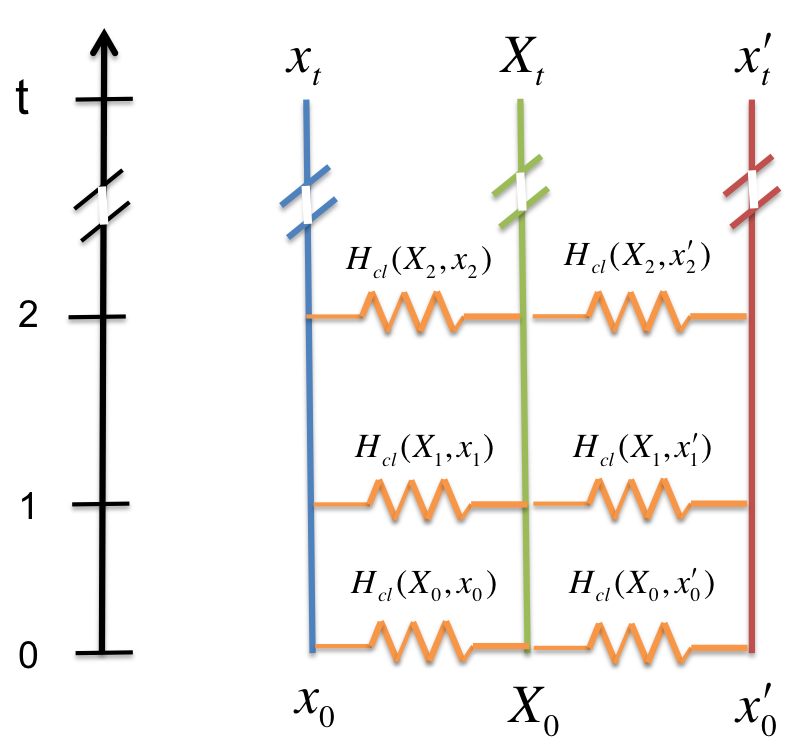}
  \caption{\label{fig1}(Color online) Schematic diagram  of the time evolution of the bath coordinates $X=(R,P)$ (the green line), the forward coherent state coordinates $x=(q,p)$ (the blue line), and the backward coherent state coordinates $x'=(q',p')$ (the red line).  The vertical axis denotes the time.  At each time step $i$, the classical Hamiltonians $H_{cl}(X_i, x_i)$ and $H_{cl}(X_i, x'_i)$ are parametrized with the updated coordinates.  The wiggly, orange lines represent the direct coupling between the evolutions of different sets of phase space coordinates under the influence of the classical Hamiltonians.
As shown, the two sets of coherent state variables are only coupled via the bath coordinates.}
\end{figure}

\end{document}